\documentclass[twocolumn]{aastex62}
\usepackage{graphicx}
\usepackage{color}
\usepackage{amsmath}
\usepackage{lineno}

\def\tcr{\textcolor{black}}

\begin{document}
\title{GW170817A as a Hierarchical Black Hole Merger}
\author{V. Gayathri}
\affiliation{Department of Physics, University of Florida, PO Box 118440, Gainesville, FL 32611-8440, USA}
\affiliation{Indian Institute of Technology Bombay, Powai, Mumbai 400 076, India}
\author{I. Bartos}
\thanks{imrebartos@ufl.edu}
\affiliation{Department of Physics, University of Florida, PO Box 118440, Gainesville, FL 32611-8440, USA}
\author{Z. Haiman}
\affiliation{Department of Astronomy, Columbia University in the City of New York, 550 W 120th St., New York, NY 10027, USA}
\author{S. Klimenko}
\affiliation{Department of Physics, University of Florida, PO Box 118440, Gainesville, FL 32611-8440, USA}
\author{B. Kocsis}
\affiliation{E\"otv\"os University, Institute of Physics, P\'azm\'any P. s. 1/A, Budapest, 1117, Hungary}
\author{S. M\'arka}
\affiliation{Department of Physics, Columbia University in the City of New York, 550 W 120th St., New York, NY 10027, USA}
\author{Y. Yang}
\affiliation{Department of Physics, University of Florida, PO Box 118440, Gainesville, FL 32611-8440, USA}

\begin{abstract}
Despite the rapidly growing number of stellar-mass binary black hole mergers discovered through gravitational waves, the origin of these binaries is still not known. In galactic centers, black holes can be brought to each others' proximity by dynamical processes, resulting in mergers. It is also possible that black holes formed in previous mergers encounter new black holes, resulting in so-called hierarchical mergers. Hierarchical events carry signatures such as higher-than usual black hole mass and spin. Here we show that the recently reported gravitational-wave candidate, GW170817A, could be the result of such a hierarchical merger. In particular, its chirp mass $\sim40$\,M$_\odot$ and effective spin of $\chi_{\rm eff}\sim0.5$ are the typically expected values from hierarchical mergers within the disks of active galactic nuclei. We find that the reconstructed parameters of GW170817A strongly favor a hierarchical merger origin over having been produced by an isolated binary origin \tcr{(with an Odds ratio of $>10^3$, after accounting for differences between the expected rates of hierarchical versus isolated mergers)}.
\end{abstract}

\section{Introduction}

During their first two observing runs, the Advanced LIGO \citep{aligo2015} and Advanced Virgo \citep{avirgo2015} gravitational-wave observatories reported the discovery of ten binary black hole mergers \citep{LIGOScientific:2018mvr}. While these events revealed considerable new information about the properties of binary black holes, it is still uncertain what astrophysical process leads to their formation and merger.

Leading possibilities for binary formation include isolated stellar binaries in which each star gives birth to a black hole through stellar core collapse, i.e. directly creating a black hole binary (hereafter isolated binaries). Alternatively, the black holes can form independently and can be brought together dynamically. This latter scenario is expected to occur in environments such as galactic centers or globular clusters with high black hole number densities (hereafter dynamical mergers). 


Within dense black hole populations it is possible that a black hole formed in a previous merger encounters new black holes and merges again \citep{2006ApJ...637..937O,2017PhRvD..95l4046G,2017ApJ...840L..24F,2019PhRvD.100d3027R,2019MNRAS.486.5008A}. This scenario, referred to as a {\it hierarchical merger}, has distinct observational signatures that makes it possible to differentiate it from other formation channels. In particular, consecutive mergers can result in heavier black holes than otherwise possible. Black holes formed during stellar core collapse are not expected to reach masses beyond about $50\,$M$_\odot$, as the heaviest stars above a critical mass explode due to pair instability without leaving a remnant behind \citep{2017ApJ...836..244W}. Therefore black holes above $\sim50\,$M$_\odot$ must have originated from something other than a single star. Hierarchical mergers present a straightforward explanation for such heavy black holes. 

\begin{figure}
   \centering  
   \includegraphics[width=0.47\textwidth]{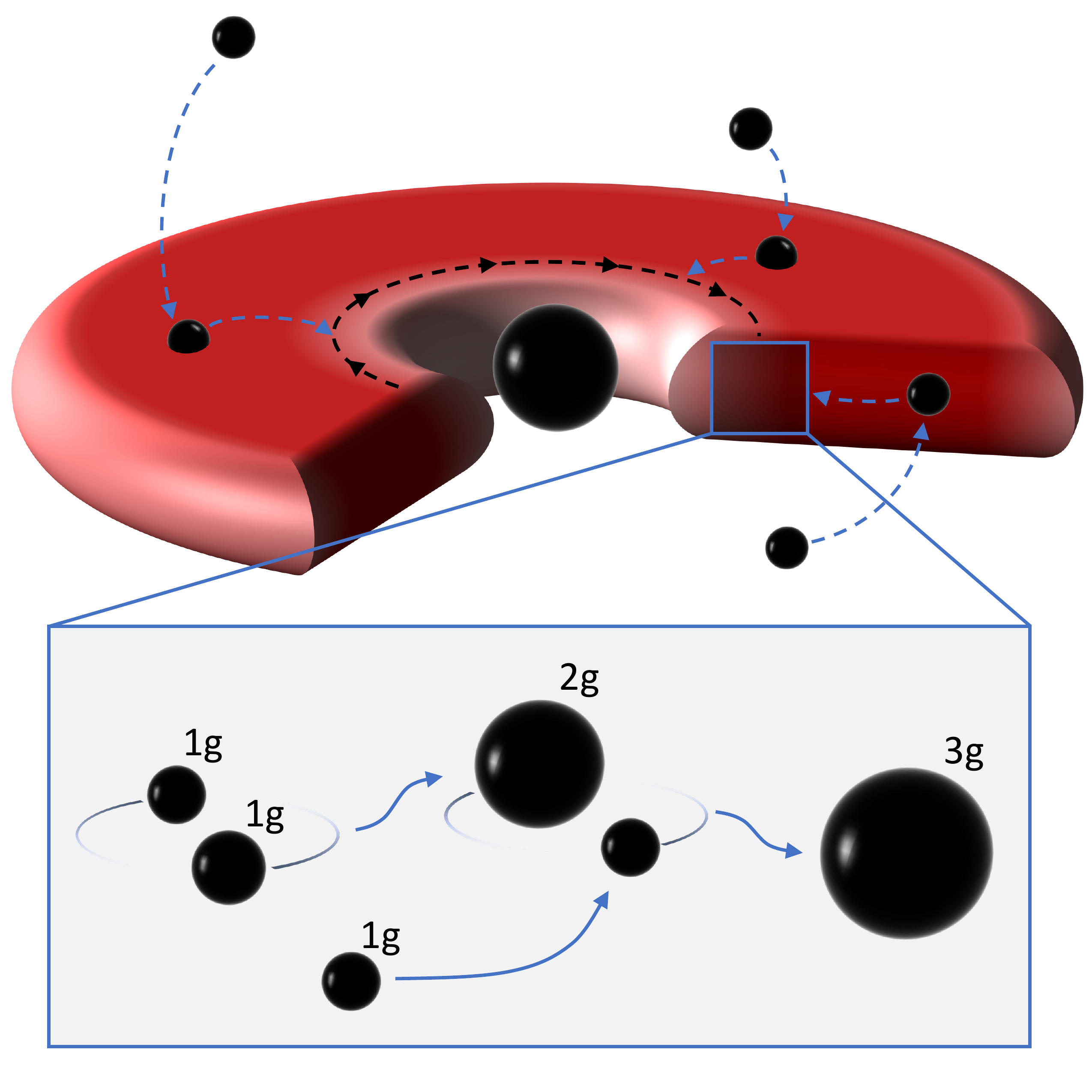}
   \caption{Illustration of hierarchical mergers in AGN disk migration traps. Stellar-mass black holes orbiting the central supermassive black holes interact with the AGN disk and gradually align their orbits with the disk's plane. Once in the disk, they experience a pressure gradient that moves them towards a migration trap in the disk, near the supermassive black hole. Black holes moving into the trap consecutively coalesce, leading to a chain of mergers.}   
\label{fig:illustration}
\end{figure}

Black hole spins can also carry the signature of hierarchical mergers. As two black holes coalesce, they form a new black hole with a characteristically high spin. In the case of the merger of two equal-mass black holes with no spin, the final black hole will be formed with dimensionless spin parameter $a\equiv cJG^{-1}M^{-2}\approx0.7$ \citep{2008ApJ...684..822B}, where $c$ is the speed of light, $J$ and $M$ are the angular momentum and mass of the black hole, respectively, and $G$ is the gravitational constant \citep{2005PhRvL..95l1101P}. The parameter is bound to be within $-1<a<1$ for any black hole. In dynamical mergers, the encounter of two black holes is typically random and the orientation of the black hole spins will be independent of the binary's orbital axis. 

An exception from this random spin orientation are mergers within Active Galactic Nuclei (AGNs). Supermassive black hole accretion disks in AGNs are expected to interact with the dense population of stellar mass black holes in the galactic center, aligning some of the black holes' orbits with the disk \citep{2012MNRAS.425..460M,2017ApJ...835..165B,2017MNRAS.464..946S,2019ApJ...876..122Y,2017NatCo...8..831B,2019MNRAS.488.4459C,2019ApJ...884L..50M}. Black holes within the disk may then migrate within the disk to so-called {\it migration traps}, due to torques exerted by the gas \citep{2016ApJ...819L..17B}. As all black hole that align their orbit with the AGN disk will end up in the migration traps, they can consecutively merge with each other; AGNs act as a black hole {\it assembly line} (see Fig. \ref{fig:illustration}). Rapid binary merger within the disk is ensured by dynamical friction, making the overall merger time $\lesssim 1$\,Myr \citep{2017ApJ...835..165B}. As black holes reside within the disk, all binary orbits will effectively be in the same plane. The spins of newly formed black holes will also be aligned (or anti-aligned) with the disk. This spin alignment can observationally distinguish hierarchical mergers within AGN disks from other hierarchical merger cases, where such alignment is atypical \citep{PhysRevLett.123.181101}. 

Looking at the 10 black hole mergers published by the LIGO Scientific Collaboration and Virgo Collaboration (hereafter LIGO-Virgo) so far \citep{LIGOScientific:2018mvr}, one of the binaries, GW170729, stands out with characteristically different features. While the black hole masses in the other 9 binaries are consistent with a power-law distribution with an upper mass cutoff around $45\,$M$_\odot$ \citep{Abbott_2019}, GW170729 appears heavier, with the mass of one of its black holes likely being above $50\,$M$_\odot$ \citep{LIGOScientific:2018mvr}. Reconstructed black hole spins are also interesting. Among spin measurements, the most accurately measured quantity through gravitational-wave observations is the binary's so-called effective spin
\begin{equation}
\chi_{\rm eff}\equiv \frac{c}{GM}\left(\frac{\vec{S}_1}{m_1} + \frac{\vec{S}_2}{m_2}\right)\cdot\frac{\vec{L}}{|\vec{L}|}
\end{equation}
where $M=m_1+m_2$ is the total mass of the binary, $m_1$ and $m_2$ are the masses of the two black holes ($m_1\geq m_2$), $\vec{S}_{1,2}$ are the spin angular momentum vectors of the black holes in the binary, and $\vec{L}$ is the orbital angular momentum vector. The measured $\chi_{\rm eff}$ for 8 of the black holes is consistent with zero, while for one event it is positive but small. In comparison, GW170729 has a higher reconstructed value, $\chi_{\rm eff}\sim0.4$. Both the mass and $\chi_{\rm eff}$ of GW170729 are consistent with a hierarchical merger occurring in an AGN \citep{2017MNRAS.471.2801S}. Nevertheless, the reconstructed parameters of GW170729 are consistent with being part of the same population as the rest of the observed black hole mergers if the prior probability of hierarchical mergers is low \citep{Abbott_2019,2019arXiv191105882F}.

With the public data release by LIGO-Virgo of their first two observing runs (O1 and O2), it became possible for external groups to carry out gravitational-wave searches. A recent such work, using a novel technique to identify signals from a single gravitational-wave detector, identified multiple possible binary black hole merger in data from the O2 observing run \citep{2019arXiv191009528Z}. The most significant of these black hole mergers, which they named GW170817A (not to confuse with the binary neutron star merger GW170817 that occurred on the same day; \citealt{2017PhRvL.119p1101A}), was reconstructed to be a binary with black hole masses and spins similar to GW170729, albeit with large uncertainties.

In this paper we examined whether GW170817A is a hierarchical merger. We compared the reconstructed parameters of this event to (i) the reconstructed distribution of LIGO-Virgo's binary black hole mergers from the O1 and O2 runs other than GW170729, (ii) the distribution expected for hierarchical mergers in AGNs \citep{PhysRevLett.123.181101}, and (iii) the distribution for a hierarchical merger scenario assuming chance encounters \citep{PhysRevD.90.104004}. 

Below we present our method in Section \ref{sec:method}, our results in Section \ref{sec:results}, and we conclude in Section \ref{sec:conclusion}.

\section{Method} \label{sec:method}

We carried out a Bayesian model comparison for GW170817A. Similar analyses have been carried out previously for other events. \cite{2017PhRvD..95l4046G} investigated whether the black hole mergers detected during LIGO's O1 observing run come from 1g or 2g mergers, and found that the data are not conclusive. \cite{2017ApJ...840L..24F} presented an analysis method of using the spin distribution of black hole mergers to probe a hierarchical merger sub-population within detected events. \tcr{In \cite{2019arXiv191104424D} proposed a self-consistent framework for generating black hole population with hierarchical merger formation and they concluded that the constrained version of proposed model is consistent with \cite{LIGOScientific:2018mvr} estimation on GWTC-1}.   \cite{Kimball:2019mfs} and \cite{2019PhRvD.100j4015C} probed whether GW170729 could originate from a 2g merger as opposed to the population derived from the rest of LIGO-Virgo's O1-O2 observations. Finally, \cite{PhysRevLett.123.181101} considered higher-generation mergers in AGN disks for explaining GW170729, in comparison with LIGO-Virgo's O1-O2 observations. Our method used here follows the prescription of \cite{2019MNRAS.486.1086M}. This is similar to the method of \cite{2017PhRvD..95l4046G}, using reconstructed masses, spins and distance. \cite{Kimball:2019mfs}, \cite{2019PhRvD.100j4015C} and \cite{PhysRevLett.123.181101} only used the reconstructed masses and spins, while \cite{2017ApJ...840L..24F} only considered the reconstructed spins.

For a given model $A$, we compute the posterior probability of $A$ given recorded gravitational data $\vec{x}$ as \citep{2019MNRAS.486.1086M}
\begin{equation}
    P(A|\vec{x}) =\pi(A) \frac{\int d \vec{\theta} P(\vec{\theta}|\vec{x})\pi(\vec{\theta})^{-1}P_{\rm pop}(\vec{\theta}|A)}{\int d\vec{\theta} P_{\rm det}(\vec{\theta})\, P_{\rm pop}(\vec{\theta}|A)}.
\end{equation}
Here, $\pi(A)\propto R_{\rm A}$ is the prior probability of model $A$, which is proportional to the expected event rate density $R_{\rm A}$ from this model.  $P(\vec{\theta}|\vec{x})$ is the probability density of true event parameters $\vec{\theta}$ given the observation. \tcr{$\pi(\vec{\theta})\propto d_{\rm L}(\vec{\theta})^2$ is the prior probability density of $\vec{\theta}$ \citep{2015PhRvL.115n1101V}, where $d_{\rm L}$ is the luminosity distance.} $P_{\rm pop}(\vec{\theta}|A)$ is the probability distribution of $\vec{\theta}$ for model $A$. $P_{\rm det}(\vec{\theta})$ is the probability of detecting a binary with parameters $\vec{\theta}$, which we take to be $1$ if the binary's distance is $<d_{\rm s}(\vec{\theta})$ and $0$ if it is $>d_{\rm s}(\vec{\theta})$ \tcr{where $d_{\rm s}$ is the sensitive distance reach.}

Considering two models, the odds ratio of model $B$ against model $A$ in explaining observations can be written as 
\begin{equation}
\mathcal{O}_{\rm B,A} = \frac{P(B|\vec{x})}{P(A|\vec{x})} = \frac{R_{\rm B}}{R_{\rm A}} K_{\rm A}^{\rm B},
\label{eq:odds}
\end{equation}
where on the right side we separated out the Bayes factor $K_{\rm A}^{\rm B}$ that is the part of the odds ratio independent of the uncertain rate densities for the different models.

In this study, we considered the following three models that are described below.

\subsection{LIGO-Virgo observation-based distribution}
This model is based on the reconstructed distribution of binary black hole parameters for all events detected by LIGO-Virgo during the O1 and O2 observing runs (model B in \citealt{Abbott_2019}). Assuming that these black holes are the end products of stellar evolution, we refer to them as 1g, or first-generation (see \citealt{2017PhRvD..95l4046G}). The primary black hole mass is distributed based on the power-law $p(m_1)\propto m_1^{-\alpha}$.  \tcr{The distributions of $\alpha$ and $m_{1(max)}$ were extracted from \cite{Abbott_2019}, which are then marginalized over following Eq. \ref{eq:odds}}. The mass distribution cuts off at a lower mass of $5\:\mbox{M}_\odot$ and upper mass limit from $m_{1(max)}$ estimation. The secondary black hole mass $m_2$ is randomly distributed within the lower mass cutoff and $m_1$. The spin amplitudes of the black holes are distributed within $0-0.9$ following a beta distribution (see Eq. 4 in \citealt{Abbott_2019}), and their orientation is randomly drawn from an isotropic distribution.

\begin{figure*}[!htp]
	\hspace{-0.5 cm}
	\centering
	\begin{minipage}[b]{.45\textwidth}
		\includegraphics[scale=0.35]{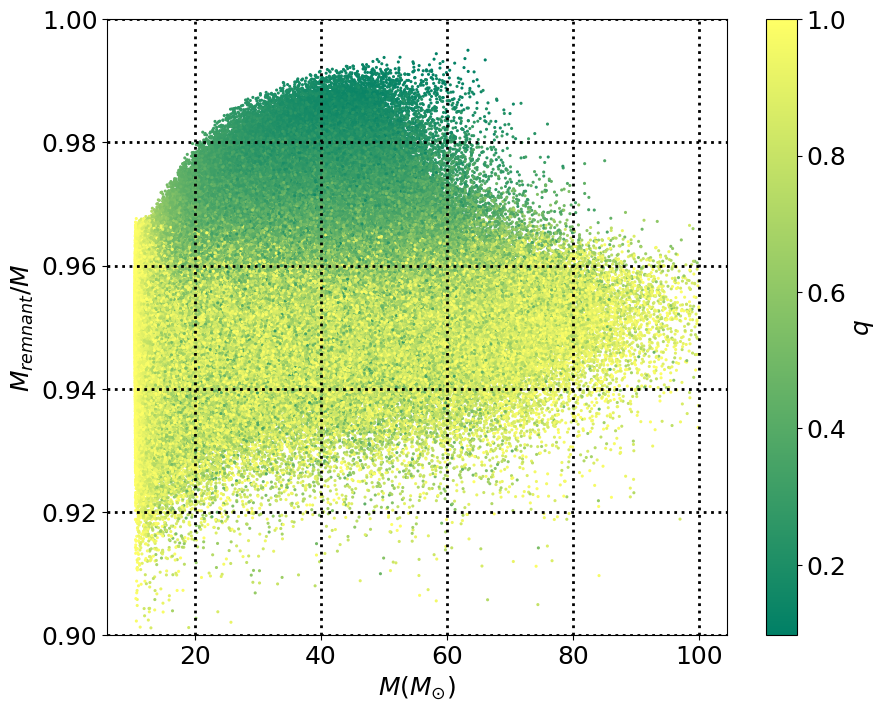}
	\end{minipage}
	\qquad 
	\begin{minipage}[b]{.45\textwidth}
		\includegraphics[scale=0.35]{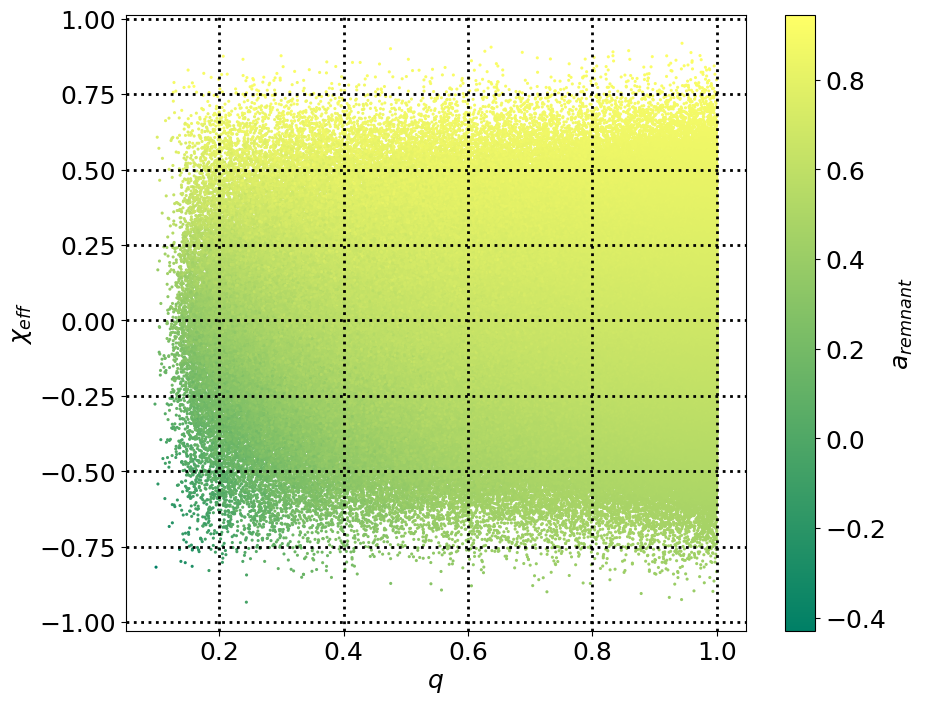}
	\end{minipage}
	\caption{Left panel:  the total mass {\it{vs}} the ratio of the remnant mass to total mass. Right panel: the mass ratio {\it{vs}} the $\chi_{\rm eff}$ for the 1g model distribution.}\label{fig:gen2}
\end{figure*}

\subsection{Hierarchical mergers in dynamical encounters (2g)}
In our second model we considered hierarchical mergers from chance encounters in dense black hole populations. We estimated the remnant mass $M_{\rm remnant}$ and spin $a_{\rm remnant}$ of a 1g merger using a higher-order phenomenological model \citep{PhysRevD.90.104004}. We drew the parameters of the initial black hole binary from the distribution of our 1g model above. We assumed that 1g black holes have zero spin. This choice does not meaningfully change the spin of the remnant black hole as spin orientations would be random in dynamical encounters. Figure \ref{fig:gen2} left and right plot shows the total mass {\it{vs}} the ratio of the remnant mass with the total mass and the mass ratio {\it{vs}} the $\chi_{\rm eff}$ for the 1g model distribution respectively. The color bar corresponds to the mass ratio of the binary black hole system. Using this information, we generated the hierarchical model distribution where primary black hole mass and spin are $M_{\rm remnant}$ and $a_{\rm remnant}$, respectively. The secondary black hole mass and spin were drawn from the distribution of our 1g model above. We note that this model neglects possible correlations between the black holes' masses, or that the merger probability depends on the black hole mass, such as in globular clusters \citep{2016ApJ...824L..12O}.  

\subsection{Hierarchical mergers in AGN disks (AGN)}
Hierarchical mergers within AGN disks have different properties than other hierarchical mergers. In our study, we adopted the hierarchical merger distribution obtained by \cite{PhysRevLett.123.181101}. This distribution was generated by simulating the orbital alignment of black holes with AGN disks, and their migration into migration traps within the disk. \cite{PhysRevLett.123.181101} found that about half the black hole mergers in AGN disks will be hierarchical, and about 20\% will be 3g or higher. As black holes individually move into migration traps, one of the black hole is every binary is 1g. 


Fig. \ref{fig:frac} shows the probability density distribution of the above three models for {\it detected binaries} over the $\mathcal{M}-|\chi_{\rm eff}|$ parameter space. For each distribution we overplotted the parameters of GW170817A as reconstructed by \cite{2019arXiv191009528Z}. We took these latter parameters using Fig. 6 of \cite{2019arXiv191009528Z}. 

\section{Results} \label{sec:results}

\begin{figure*}
   \centering  
   \includegraphics[width=\textwidth]{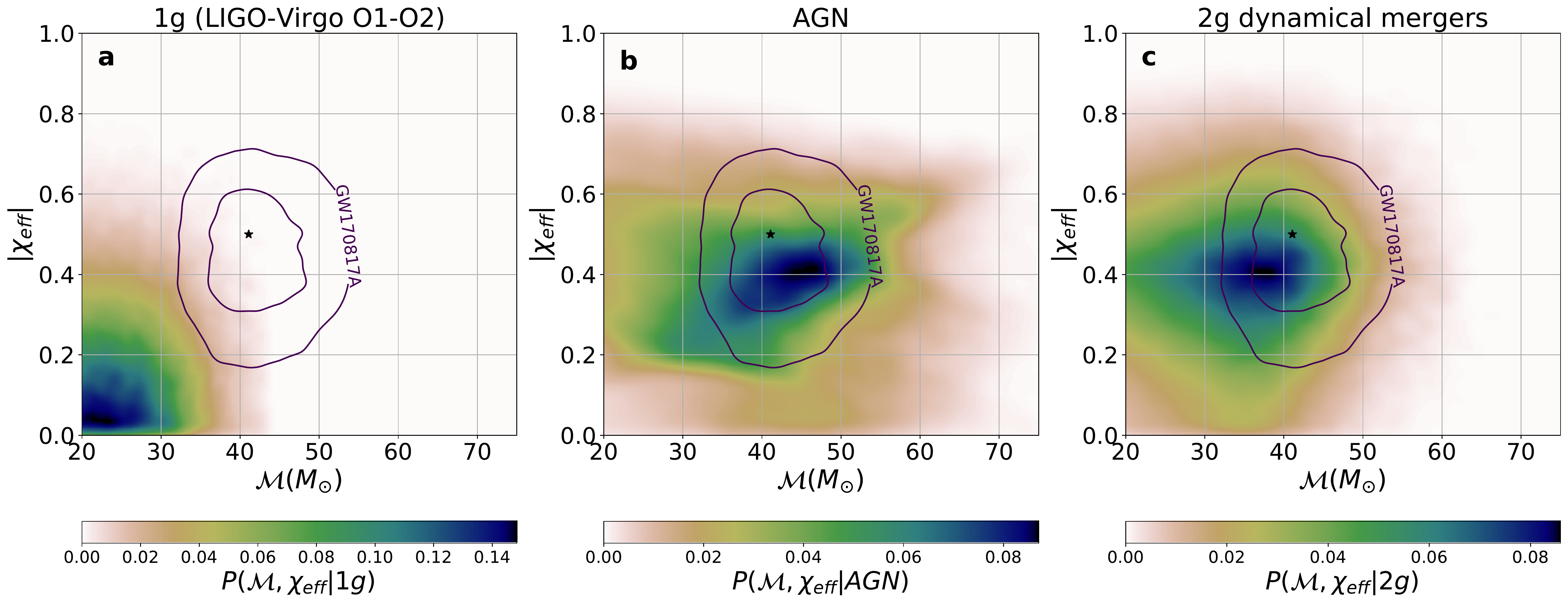}
   \caption{Expected probability density of the chirp mass $\mathcal{M}$ and effective spin $\chi_{\rm eff}$ for detected gravitational-wave events, for different underlying population models: (a) distribution based on LIGO-Virgo's O1-O2 detections (Model B in \citealt{Abbott_2019}); (b) expected hierarchical merger distribution in AGN disks \citep{PhysRevLett.123.181101}; (c) 2g hierarchical mergers assuming no spin alignment. For comparison, we show the reconstructed parameters of GW170817A \citep{2019arXiv191009528Z}.}   
\label{fig:frac}
\end{figure*}

We computed the Bayes factor for the three possible combinations of our three models using Eq. \ref{eq:odds} with $\vec{\theta}=\{m_1,m_2,\chi_{eff}\}$. We obtained $K^{\rm AGN}_{\rm 1g}\approx 2 \times 10^4$, $K^{\rm 2g}_{\rm 1g}\approx 3 \times 10^4$, and $K^{\rm AGN}_{\rm 2g}\approx 0.7$. 

We see that a hierarchical merger origin better explains the observation for GW170817A than a 1g merger. As shown in Fig. \ref{fig:frac}, this is due to both the expected higher mass and higher effective spin of hierarchical binaries compared to LIGO-Virgo's O1-O2 detections (other than GW170729). We also see that an AGN-based hierarchical merger is a marginally less likely explanation of GW170817A than a 2g merger from a dynamical encounter.

Determining the origin of GW170817A depends on not just the above Bayes factors, but also the expected rates of the different formation channels. The rate density of binary black holes is known from LIGO-Virgo observations to be $\sim10-100$\,Gpc$^{-3}$\,yr$^{-1}$ \citep{LIGOScientific:2018mvr}\footnote{The quoted rate interval is for mixture models based on \cite{LIGOScientific:2018mvr}. Population-based models give a somewhat higher range with lower bound $\gtrsim30$\,Gpc$^{-3}$\,yr$^{-1}$ \citep{Abbott_2019}. The odds ratios quoted below adopt the $\sim10-100$\,Gpc$^{-3}$\,yr$^{-1}$. Using the rate estimate of population based models would result in a somewhat increased likelihood of the 1g model.}. The situation is less clear for hierarchical mergers. 

\cite{2019ApJ...876..122Y} estimate that the mergers within AGNs is $\sim4$\,Gpc$^{-3}$yr$^{-1}$ (see also \cite{2019arXiv191208218T} who find a broader range of possible merger rate densities within $0.02-60$\,Gpc$^{-3}$yr$^{-1}$). While the fractional merger rate in AGNs is relatively lower, the spectral hardening during orbital alignment \citep{2019ApJ...876..122Y} and the hierarchical process increases the black hole masses compared to the black holes' initial mass distribution, resulting in greater detection volume for LIGO-Virgo. Using the above merger rate density estimates, the odds ratio of the AGN model vs. the 1g LIGO-Virgo population is $\mathcal{O}^{\rm AGN}_{\rm 1g}=10^3-10^4$. 

To estimate the odds ratios for our 2g model from dynamical encounters, we consider recent Monte Carlo simulations of globular clusters coupled with their intragalactic evolution, which showed that the rate of all black hole mergers from globular clusters is in the range of $4-60\,\rm Gpc^{-3} yr^{-1}$ at redshift $z<0.5$ \citep{2016PhRvD..93h4029R,2018PhRvL.121p1103F,2018ApJ...866L...5R}. The rate of 2g mergers among these mergers is estimated to be $10\%$ of the total rate density \citep{2016ApJ...824L..12O,2019PhRvD.100d3027R}. As the predicted merger rate of $60\,\rm Gpc^{-3} yr^{-1}$ is higher than part of the allowed range of overall merger rate measured by LIGO-Virgo, for the odds ratio calculation we limit the globular-cluster rate to less than or equal to the overall LIGO-Virgo rate. The corresponding odds ratios are $\mathcal{O}^{\rm 2g}_{\rm 1g}>100$. We only have a lower limit here as, based on the quoted number above, it is possible that all LIGO observations come from globular clusters. Comparing the two hierarchical channels, we obtain $\mathcal{O}^{\rm AGN}_{\rm 2g}=0.4-2$. 

\section{Discussion} \label{sec:discussion}

Our 2g model only includes second-generation mergers. Hierarchical mergers from dynamical encounters could also lead to higher-generation mergers, which would lead to higher typical masses. However, the typical effective spin would not significantly increase as the spin orientation in dynamical encounters is random. Therefore, we do not expect our likelihood for GW170817A to significantly change if we include 3g+ mergers in our 2g dynamical encounter model.

Besides hierarchical mergers that we considered here, we note that accretion onto black holes can also increase their mass and effective spin, similarly to the properties of GW170817A. Such a scenario may occur in dense environments such as AGN disks \citep{Yi_2019}, or due to significant fallback accretion \citep{2019arXiv190704218P}.

The selection of GW170817A for this analysis was based on its high observed mass and spin. In order to make a quantitative statement about the observed binary black hole population as a whole, a comprehensive study of all detected events will be necessary.

\section{Conclusion} \label{sec:conclusion}

We examined whether the newly identified binary black hole merger, GW170817A, could be a hierarchical merger. While the event is not as certain to be astrophysical as some of LIGO-Virgo's other discoveries, assuming it is a gravitational-wave signal we found that the event is significantly more likely to have been a hierarchical merger compared to coming from the same population as LIGO-Virgo's O1-O2 events published in the GWTC-1 catalog. 

In particular, a hierarchical merger in an AGN disk is expected to have a chirp mass and an effective spin centered around $\mathcal{M}\sim 45\,$M$_{\odot}$ and $\chi_{\rm eff}\sim0.4$, close to the reconstructed parameters of GW170817A, i.e. $\mathcal{M}\sim 40\,$M$_{\odot}$ and $\chi_{\rm eff}\sim0.5$. We obtained a Bayes factor $K^{\rm AGN}_{\rm 1g}\approx2\times10^4$ comparing this AGN model to LIGO-Virgo's O1-O2 detections, corresponding to an odds ratio of $\mathcal{O}^{\rm AGN}_{\rm 1g}=10^3-10^4$. 

GW170817A is also consistent with a hierarchical merger from dynamical encounters. We found that the Bayes factor of a 2g merger over mergers drawn from LIGO-Virgo's O1-O2 distribution is $K^{\rm 2g}_{\rm 1g}\gtrsim100$.

Together with GW170729 which had very similar reconstructed parameters ($\mathcal{M}\sim 35\,$M$_{\odot}$ and $\chi_{\rm eff}\sim0.4$), GW170817A may be the first example for an exciting new population of hierarchical black hole mergers. With further similar observations it is possible that they will emerge as the first definitively identified source of origin for black hole mergers.

\section{Acknowledgment}
The authors thank Doga Veske and Tito Dal Canton for useful feedback. The authors are thankful to the University of Florida and Columbia University in the City of New York for their generous support. The Columbia Experimental Gravity group is grateful for the generous support of the National Science Foundation under grant PHY-1708028.  VG acknowledges Inspire division, Department of Science and Technology, Government of India for the fellowship support. This project was supported by funds from the European Research Council (ERC) under the European Union's Horizon 2020 research and innovation programme under grant agreement No 638435 (GalNUC) and by the Hungarian National Research, Development, and Innovation Office grant NKFIH KH-125675 (to BK). ZH acknowledges support from NASA grant NNX15AB19G and NSF grant 1715661.
\bibliography{Refs}
\end{document}